\documentclass[prl,twocolumn,superscriptaddress,showpacs,preprintnumbers,amsmath,amssymb]{revtex4}

\usepackage{graphicx}
\usepackage{dcolumn}
\usepackage{bm}
\usepackage{epsfig}
\newcommand{\ignore}[1]{}

\begin{document}

\title{Quantum Dot in Z-shaped Graphene Nanoribbon}

\author{Z. F. Wang}
\address{Hefei National Laboratory for Physical Sciences at Microscale}%
\author{Huaixiu Zheng}
\address{Electrical and Computer Engineering, University of Alberta, AB T6G 2V4, Canada}

\author{Q. W. Shi}
\email[Author to whom correspondence should be addressed. Electronic address:]{phsqw@ustc.edu.cn}
\address{Hefei National Laboratory for Physical Sciences at Microscale}%

\author{Jie Chen}
\email[Author to whom correspondence should be addressed. Electronic address:]{jchen@ece.ualberta.ca}
\address{Electrical and Computer Engineering, University of Alberta, AB T6G 2V4, Canada}%
\address{National Institute of Nanotechnology, Edmonton AB T6G 2V4, Canada }%

\author{Qunxiang Li}
\address{Hefei National Laboratory for Physical Sciences at Microscale}%

\author{J. G. Hou}
\address{Hefei National Laboratory for Physical Sciences at Microscale}%

\date{\today}

\begin{abstract}
Stimulated by recent advances in isolating graphene, we discovered that
quantum dot can be trapped in Z-shaped graphene
nanoribbon junciton. The topological structure of the junction can confine electronic
states completely. By varying junction length, we can alter the spatial confinement
and the number of discrete levels within the junction. In addition,
quantum dot can be realized regardless of substrate induced static disorder or irregular edges
of the junction. This device can be used to easily design quantum dot devices.
This platform can also be used to design zero-dimensional
functional nanoscale electronic devices using graphene ribbons.
\end{abstract}

\maketitle

Due to its amazing electrical properties, carbon nanotubes (CNTs)
discovered by Iijima \cite{1} in 1991, have been considered as a
leading candidate for nanoscale electronic applications.
Major experimental and theoretical breakthroughs have been achieved
\cite{2,3,4}, combining two distinct chirality carbon-nanotubes by
introducing topological point defects in the graphene hexagonal
lattice, to realize the quantum dot in the carbon nanotube
heterojunctions. In such devices, the major sources of spin
decoherence have been identified as the spin orbit interaction,
coupling the spin to lattice vibrations and the hyperfine interaction
of the electron spin with the surrounding nuclear spins. Therefore,
it is desirable to form qubits in quantum dots based on these materials,
where spin orbit coupling and hyperfine interaction are considerably
weaker \cite{5}. However, large scale integrated nanotube quantum dot
devices are hard to make, because it is still difficult to assemble
large numbers of CNTs together.

Very recently, the fabrication of a single layer of graphene and the
measurement of its electric transport properties have been achieved \cite{6,7}.
Graphene is a flat monolayer of carbon atoms tightly packed into a
two-dimensional (2D) honeycomb lattice. What makes graphene so attractive
for nanoelectronics is that the energy spectrum closely resembles the
Dirac spectrum for massless fermions. For massless Dirac fermions, the
band gap is zero and the linear dispersion law holds at low energy,
quasiparticles in graphene behave differently from those in conventional
metals and semiconductors, where the energy spectrum can be approximated
by a parabolic dispersion relation. Research interest in this material
has grown exponentially \cite{6,7,8,wzf}.

\begin{figure}[htpb]
\begin{center}
\epsfig{figure=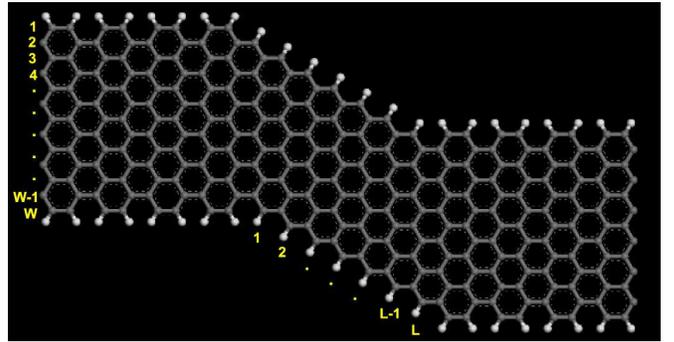,height=4.5cm}
\end{center}
\caption{(color online) Atomic structure of quantum-dot device
made of perfect Z-shaped GNR junction. Width of the armchair GNR
is labeled by integer W. Length of junction is labeled by integer L. }
\end{figure}

As the building block of the carbon nanotube (CNTs), graphene can be
viewed as a sheet of a unrolled single-walled nanotube. Graphene has
similar mechanical, thermal and electrical properties as those of
CNTs. Unlike CNTs, its flat structure can be easily etched using
conventional lithography techniques. The interconnection wires become
unnecessary and integrated nanoelectronic devices can be completely
made by using continuous graphene sheets. Armchair and zigzag
graphene nanoribbons (GNR) are two basic graphene ribbon structures.
Based on the nearest neighbour $\pi$ orbital tight binding model with
diagonal matrix element fixed at Fermi level $\epsilon_F=0 eV$ and all nonzero
off diagonal matrix element set to $\gamma= -2.66 eV$, zigzag GNRs are
always metallic and the armchair GNRs are either metallic or semiconducting
depending on its width $W$ \cite{9}. For instance, when $W=3n-1$, it is
metallic, otherwise is semiconducting. Comparing with the recent Local density
approximation (LDA) results \cite{10,11}, except for the band gap opening for
the ribbon with narrow width as the consequence of $\sigma$ bond length changing,
GNR's electronic structure can still be quantitatively described using the simple
tight binding model.

In addition to the study of 2-D and 1-D electronic proprieties of graphene,
research attention has recently been focus on designing quantum dot (the 0-D device)
base on this novel material \cite{12,13,14,dot}. In this letter, we propose a GNR
quantum dot device, which consists of a Z shaped zigzag GNR junction connecting to
two semi-infinite armchair GNRs as shown in Fig.1. Here, we suppose that GNR edge
bonds are saturated by hydrogen atoms and no distortions exist in these GNRs.
According to our numerical results, we find that this Z-shaped junction
device can completely confine electronic states induced by the topological
structure of the junction. By varying the length of the junction, the spatial
confinement and the number of discrete levels are modified accordingly.
Surprisingly, these confined states can still exist even when considerable
static disorders and irregular edges of the junction occur. This finding
show that this quantum-dot device can be made without too many constraints,
which indicates that such a device can be easily fabricated.

To study the electronic proprieties of the Z-shaped GNR junction,
we separate the device as shown in Fig. 1 into three regions:
the left lead, the middle junction, and the right lead.
In this study, we assume that the junction width is $W-1$, the left/right
leads have equal width $W$ and the length of the junction is $L$. Here
$L$ and $W$ are integer. In this design, the leads are semi-infinite
armchair GNRs, but the junction is a zigzag GNR.
We have performed calculations using the nearest neighbour $\pi$
orbital tight binding model, the density of states (DOS)
of the Z junction are determined by the direct diagonalization of
$H=H_c+\Sigma_L^{r}+\Sigma_R^{r}$, where $H_c$ is the Hamilton
of Z junction. Including parts of the armchair GNRs in the Hamilton
of $H_c$ do not obviously change our results, so in the following
calculation, only the zigzag junction part is included in the $H_c$.
The contributions from the left and right leads are included in
the self-energy term $\Sigma_{L/R}^{r}$,
which are calculated using Green's function along with a transfer
matrix technique \cite{15,16}. The DOS of this system can be expressed
as $DOS(E)=\sum\limits_\alpha\frac{1}{2\pi}\frac{\gamma_\alpha}
{(E-\varepsilon_\alpha)^2+(\gamma_\alpha)^2}$.
The summation covers all eigenvalues. $\varepsilon_\alpha$
is the real part of eigenvalue and represents the position of the
state. $\gamma_\alpha$ is the imaginary part of the eigenvalue,
which represents the broaden of the state. Local density of states
(LDOS) are directly obtained from Green's function.

\begin{figure}[htpb]
\begin{center}
\epsfig{figure=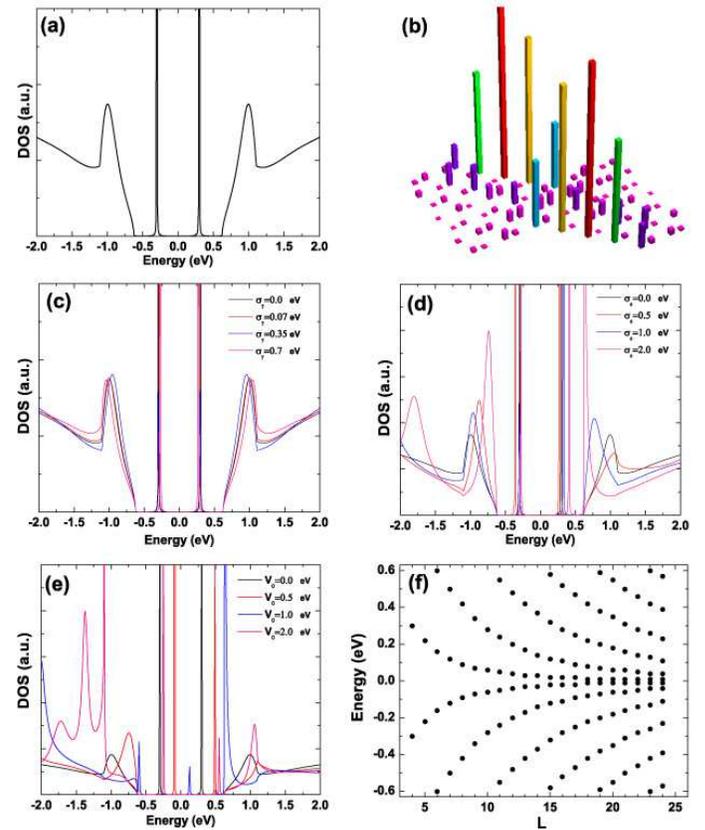,height=11cm}
\caption{(color online) Z-shaped junction with width $W=7$ and length $L=4$.
(a) DOS of perfect junction without disorder. (b) The corresponding
spacial distribution of LDOS of the perfect junction for E=-0.3 eV. (c)
DOS of the junction when present short range disorder, $\sigma_\gamma$
changes with 0, 0.07, 0.35 and 0.7 eV. (d) DOS of the junction when
present short range disorder, $\sigma_\epsilon$ changes with 0, 0.5, 1.0
and 2.0 eV. (e) DOS of the junction when present long range disorder with
one centered Gaussian potential $V(r)=V_0exp(-r^2/2\sigma^2)$, sites on
the lattice, here $\sigma=3.5$ {\AA} and $V_0$ changes with 0, 0.5, 1.0
and 2.0 eV. (f) Length dependence for the energy of confined states.}
\end{center}
\end{figure}

In Fig.2 (a), we show the DOS of a Z-shaped junction with $W=7$ and
$L=4$. There are two sharp peaks in the energy range within the band
gap. The position of the peaks is symmetric around $E=0 eV$, because
the topological structure of the junction does not break electron-hole
symmetry. This phenomenon is dramatically different from the quantum dot
in nanotube heterojunctions, in which the pentagon-heptagon interface
break down this symmetry \cite{2}. In Fig.2 (b), we show the spatial
dependence of LDOS for the discrete states in a Z-shaped junction. In
this case, the discrete energies are $E_1= -0.3 eV$ and $E_2= 0.3 eV$.
By considering energy symmetry, only the LDOS of $E_1= -0.3 eV$ is
plotted in Fig.2 (b). Since the LDOS corresponds to the squared
amplitude of the wave function, Fig.2 (b) illustrates the spatial
localization of the Z shaped junction, i.e. the discrete states are
localized in the junction region and especially along the edge of
the junction. This phenomenon clearly demonstrates that these discrete
states are caused by quantum confinement. The Z-junction design can be
used as a quantum dot device. Unlike previous designs \cite{12,13,14},
quantum confinement in our study is due to the change in the network
connectivity. These confined states are attributed to the surrounding
barriers, which form by the interconnection between the armchair GNR
and the zigzag GNR, which has been well used to explained the nanotube
heterojunctions quantum dot \cite{2,3}.

Perfect GNRs seldom exists in reality. Even if initial perfect,
once physical adsorbed on a surface, GNRs will experience the effects
of the disorder due to the interaction with the substrate \cite{17}.
This disorder can change their wave function and affect their usage in
nanoelectronic devices. According to the potential range of the disorder,
they can be classified into long-range and short-range effects.
The long-range effect (potential changes longer than the lattice distance) is
caused by substrate charge and the short-range effect (potential varies
rapidly at the scale of the lattice distance) is caused by residual interaction.
The effect of these disorders can be simulated by adjusting corresponding elements
in the Hamilton matrix \cite{17}. For instance, the short-range disorder effect can be
modeled by assuming that the diagonal and the nonezero off diagonal matrix elements
independently fluctuate around their initial values with variances $\sigma_\epsilon$
for diagonal elements and $\sigma_\gamma$ for nonzero off diagonal elements.
The long range disorder can be stimulated by introducing a 2D Gaussian potential with
the form of $V(r)=V_0exp(-r^2/2\sigma^2)$ centered around the carbon site
to shift the diagonal element. In the following calculation, the disorder is
limited in the junction region for simplicity.

The effects of the substrate induced static disorder to the confined
states in the Z-shaped junction are shown in Fig.2 (c),(d) and (e).
To see the disorder effects more clearly, we consider only one type of
disorders in Fig.2 (c) $\sim$ (e), respectively. In Fig.2 (c), we suppose
that only the short-range disorder with $\sigma_\gamma$ exists within
the junction region. Considering the C-C bond of length $d$, the corresponding
nonzero off diagonal element $\gamma$ will change $\delta\gamma=\alpha\delta d$
with $\alpha\simeq47 eV/nm$ \cite{18}. So in Fig. 2(c), for
$\sigma_\gamma= 0.07, 0.35$ and $0.7 eV$, the C-C bond lengths change as much as
$\pm1 \%, \pm5 \%$ and $\pm10 \%$. Even the bond length changes  $10 \%$.
The DOS, however ,does not change much except for a slight peak shift of the
discrete states. This off-diagonal short-range disorder has almost no impact
to the confined states. In addition, the DOS remains symmetric around $E=0 eV$,
which indicates that this type of disorder cannot break the electron-hole symmetry
in this junction device. Secondly, in Fig.2 (d), we only consider the short-range
disorder $\sigma_\epsilon$. With different disorder strengths as much as
$\sigma_\epsilon= 0.5, 1.0$ and $2.0 eV$, fortunately, these confined states can
still exist. Unlike the former case, the position of the confined states change
more dramatically for the $\sigma_\epsilon$ disorder and they are not no longer
symmetric around $E=0 eV$. Lastly, for the long-range disorder, a Gaussian potential
is introduced in the junction range. The shift of different potential center,
however, will not change our findings (we only plot one of them in Fig.2 (d) for
illustration). Compare to the $\sigma_\epsilon$ disorder, the DOS becomes more
asymmetric around $E=0 eV$. Two confined states, however, still exist within the
gap region. These phenomena can be easily explained, because the discrete states are
confined by the two barriers formed by the change in the network connectivity.
The substrate induced disorder cannot change the topological barriers much, and
thus it still has enough strength to confine the electron in the junction and
form a quantum dot.

\begin{figure}[htpb]
\begin{center}
\epsfig{figure=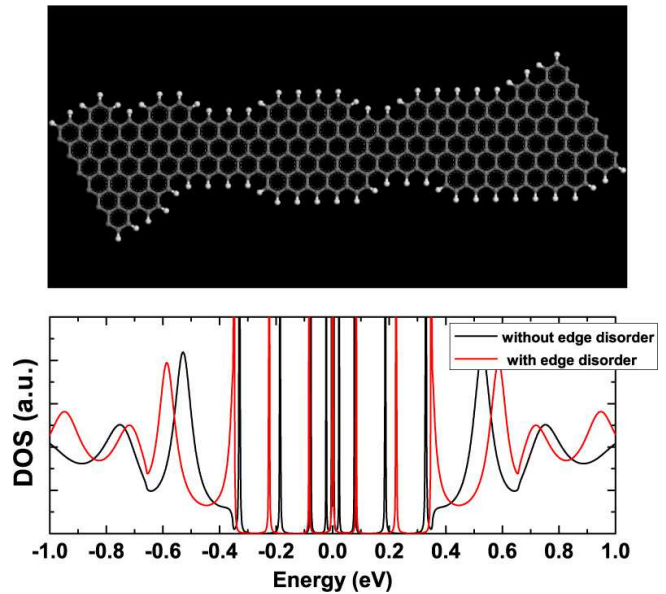,height=7.8cm}
\caption{(color online) Top figure: the atomic structure of a
Z-shaped junction of $W=13$ and $L=20$ with irregular edge.
Bottom figure: The DOS of a perfect Z junciton and an irregular edge
Z junction.}
\end{center}
\end{figure}

To extend this novel quantum-dot design, we also investigate
how the junction length affects quantum-dot confinement. The length
of the quantum-dot along the junction is $L \times 2.13{\AA}$. Here we
assume no disorder exists in the junction region, Fig.2 (f) shows the
energy of the confined states vs. the length of the quantum-dot. As
expected, when $L$ increases, the number of confined states increases
as well. Furthermore, with the increasing length of this quantum-dot
device, the confined states around $E=0 eV$ are almost degenerated.
In addition, within in the range of the L value studied,  the energy
spacing between the discrete level is about $\simeq100 meV$ except for
these states around $E=0 eV$. This value is larger than the thermal
broadening at room temperature.

Unlike CNT devices, GNR device most likely have irregular edges due to
lithography. Finally in our study, we consider the impact of edge
disorder in zigzag junction region. Here, we follows the method suggested
by Areshkin to simulate eroded zigzag edge \cite{17}. This method ensure
that all the C atoms remains 3-fold coordinated with at most one H atom
attached to an edge C atom without steric problem. Here we restrict the
erosion to the outmost layer of the edge. Fig.3 shows how these defects
affect the confined states. Similar to the case of the $\sigma_\gamma$
disorder, a slight peak shift can be observed. However, the confined states
still exist even under the presence of the irregular edges. We can explain it
in a straight way. This edge disorder erodes some carbon and hydrogen atoms
in the zigzag junction. The effects of breaking bonds are equivalent to add
$-\gamma$ to the nonzero off diagonal elements in the Hamilton matrix similar
to the $\sigma_\gamma$ disorder. The effects for eroded hydrogen can be neglected
because carbon atoms can form strong covalent bonds with hydrogen atom.
The $\sigma$ bands of these bonds lie far away from the Fermi surface and do not
need to be considered. This result is extremely important and provides more
convenience and flexibility to implement this structure in experiments.
With the increasing width of the GNR junction, the ribbon behaviors will gradually
transit to that in a 2D graphene. The edge effects would become much weaker.
Therefore, a wider GNR junction should be able to tolerate even a greater degree of
edge disorder. However, these imperfections do not change the behavior of
our quantum-dot device, which is very important in future nanofabrication.

In summary, we propose a quantum-dot device design using
a Z-shaped GNR junction. This system can completely confine electronic
states induced by the topological structure of the Z-junction. By
varying the length of this junction device, the spatial confinement
and the number of discrete levels can be modified. In addition, the
substrate induced static disorder and the irregular edges of GNRs
do not destroy these confined states. These findings provide a
convenient way to fabricate this structure experimentally.
For our on-going study, we are investigating the effect of coulomb
interactions in this structure. Overall, our contribution is to provide
a simple model to design zero-dimensional functional nanoscale electronic
devices using graphene ribbons.

This work is partially supported by the National Natural Science
Foundation of China with grant numbers 10574119, 10674121 and
50121202. The research is also supported by National Key Basic Research
Program under Grant No. 2006CB922000, Jie Chen would like to acknowledge
the funding support from
the Discovery program of Natural Sciences and Engineering Research
Council of Canada under Grant No. 245680.

\end{document}